\documentclass[aps,prl,twocolumn,groupedaddress]{revtex4}
\usepackage{amssymb,amsmath,amsfonts,graphicx,latexsym,bm,color}
\usepackage{hyperref}

\newcommand{\beq}{\begin{eqnarray}}
\newcommand{\eeq}{\end{eqnarray}}
\newcommand{\bem}{\begin{pmatrix}}
\newcommand{\eem}{\end{pmatrix}}

\newcommand{\hb}{\hbar}

\newcommand{\f}{\frac}

\newcommand{\tb}[1]{\textbf{#1}}
\newcommand{\tr}[1]{\textrm{#1}}
\newcommand{\ro}[1]{\sqrt{#1}}

\def\e{\epsilon}
\def\d{\delta}
\def\D{\Delta}

\def\p{\phi}

\def\s{\sigma}

\def\lt{\left}
\def\rt{\right}

\def\da{\downarrow}
\def\ua{\uparrow}

\begin{document}
\title{Strongly Interacting Two-Dimensional Dirac Fermions}
\author{Lih-King Lim$^{1}$, Achilleas Lazarides$^{1}$, Andreas Hemmerich$^{2}$, and C. Morais Smith$^{1}$}
\affiliation{$^{1}$Institute for Theoretical Physics, Utrecht University, Leuvenlaan 4, 3584 CE Utrecht, The Netherlands}
\affiliation{$^{2}$Institut f\"{u}r Laser-Physik, Universit\"{a}t Hamburg, Luruper Chaussee 149, 22761 Hamburg, Germany}
\date{\today}

\begin{abstract}
We show how strongly interacting two-dimensional Dirac fermions can be realized with ultracold atoms in a two-dimensional optical square lattice with an experimentally realistic, inherent gauge field, which breaks time-reversal and inversion symmetries. We find remarkable phenomena in a temperature range around a tenth of the Fermi-temperature, accessible with present experimental techniques: at zero chemical potential, besides a  conventional $s$-wave superconducting phase, unconventional superconductivity with non-local bond pairing arises. In a temperature versus doping phase diagram, the unconventional superconducting phase exhibits a dome structure, reminiscent of the phase diagram for high-temperature superconductors and heavy fermions.
 \end{abstract}
\maketitle
\date{\today}

Some of the most intriguing low temperature phenomena in solids occur due to strong correlations in low-dimensional electron systems. Prominent examples are the high-temperature superconductivity in the cuprates, the unconventional superconductivity in heavy-fermion compounds, or the fractional quantum Hall effect. An exciting novel aspect to the physics of electronic matter arises when the electrons behave like Dirac fermions. The significance of Dirac-like dispersion for understanding strongly correlated electronic systems is not well known. The recent breakthrough in fabricating sheets of graphene \cite{Neto:09} has provided us with a laboratory model of two-dimensional Dirac fermions in a hexagonal crystal lattice. The preparation of a well controlled, clean system of strongly interacting Dirac fermions in a solid state context appears intricate, if not impossible. In contrast to solid state systems, ultracold atoms in optical lattices can be controlled to perfection in experiments, which has allowed for experimental demonstrations of prototypical many-body phenomena, as the superfluid-Mott insulator transition in the Bose-Hubbard model \cite{Jaksch:98, Greiner:02, Spielman:08}. 

\begin{figure}[t]
\includegraphics[scale=.24, angle=0, origin=c]{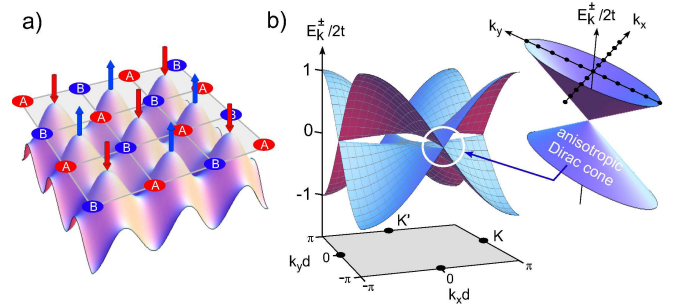}
\caption{\label{Fig.1} (color online) a) A staggered gauge field (arrows), induced by a time-dependent optical potential, splits the lattice into two sublattices $\mathbb{A}$ and $\mathbb{B}$ \cite{Hem:07,Lim:08}. In the tight binding regime, the particles hop between the potential minima and pick up a flux $\pm\p$ around a unit plaquette. b) The single-particle energy spectrum $E_{\tb{k}}^{\pm}$ in the first Brillouin zone of the Hamiltonian~(\ref{Ham1}). The two inequivalent Dirac points are indicated $\tb{K}$ and $\tb{K}'$. }
\end{figure}

In this Letter, we show how a {\it square} optical lattice can be used to realize {\it strongly interacting two-dimensional Dirac fermions}. We employ an experimentally realistic optical square lattice \cite{Hem:07} with an inherent effective staggered gauge field, recently shown to provide a Dirac-like single-particle spectrum at the Brillouin zone edge \cite{Lim:08} (see Fig.~\ref{Fig.1}a). Notably, this lattice geometry breaks time-reversal and inversion symmetries. We consider a mixture of condensed bosons and a balanced mixture of spin-up and spin-down fermions subjected to this lattice. The induced on-site and nearest-neighbor attractive interactions for the fermions are shown to be independently tunable, thus allowing for the study of the competition between different strongly correlated BCS ground states. For temperatures around a tenth of the Fermi-temperature $T_F$, accessible with state of the art experimental cooling techniques, we find robust exotic phenomena: in addition to the conventional local-pairing $s$-wave superconductivity, an unconventional superconductivity with non-local bond pairing arises, which spontaneously breaks both, the $U(1)$ gauge and the $C_{4v}$ crystal lattice symmetries. On the other hand, by varying the fermion filling fraction at a fixed coupling strength, the phase diagram exhibits a superconducting dome surrounded by a Dirac-liquid and a Fermi-liquid at low and high doping, respectively. Exploring its details in a well controlled experimental environment may yield new insights into the phenomenology of high-$T_c$ superconductors or heavy fermion systems, which exhibit strikingly similar phase diagrams.

\textit{Tailoring the Strongly Interacting Dirac Fermions}:\label{Intera}
In contrast to other recently proposed schemes to achieve Dirac-like single particle spectra in optical lattices \cite{Liu:09, Goldman:09}, we employ a square lattice with an effective staggered magnetic field (see Fig.~\ref{Fig.1}a) \cite{Hem:07,Lim:08}. In the tight-binding limit, the kinetic energy describing tunneling in this lattice takes the form
\beq\label{Ham1}
H_0= -t\sum_{<ij>}e^{i \p_{ij}/4} a^\dag_i b_j+\tr{h.c.}
\eeq
Here,  $t$ is the hopping amplitude and $\p_{ij}=\pm \p$ is the anisotropic phase factor (with the sign depending on the indices $i$ and $j$), which denotes the presence of a dimensionless magnetic flux $\p$ (in units of the fundamental flux quantum) alternating in sign across the plaquettes. The operators $a^\dagger_i$  $(a_i)$ and $b^\dagger_j$  $(b_j)$ create (destroy) a particle in the sublattices $\mathbb{A}$ and $\mathbb{B}$, respectively (see Fig.~\ref{Fig.1}a). They satisfy canonical commutation (anti-commutation) relations for bosonic (fermionic) particles. The spacing between adjacent $\mathbb{A}$ sites is $d \equiv \lambda/ \sqrt{2}$, with $\lambda$ being the wavelength of the optical lattice.

For ultracold single-component fermionic atoms, interactions are absent due to negligible $s$-wave scattering and the system is described by the kinetic Hamiltonian in Eq.\ (\ref{Ham1}). At unit filling fraction, the low-energy band structure possesses two inequivalent  {\it anisotropic} Dirac cones at $\tb{K}=(\pi/d,0)$ and $\tb{K}'=(0,\pi/d)$ along the Brillouin zone edges  (see Fig.~\ref{Fig.1}b), with isotropy recovered for $\p=\pi$. Around $\tb{K}$, the Hamiltonian is given by
\beq\label{Dirac}
H_{0,F}=\sum_{\tb{k}} \alpha\left[\cos\lt(\frac{\p}{4}\rt)  k_x- i\sin\lt(\frac{\p}{4}\rt)  k_y \right]a^{\dag}_{\tb{k}}b_{\tb{k}}+\tr{h.c.},
\eeq
where $\alpha=\ro{2}\hb v_{F}$, $v_{F}=\ro{2}t d/\hb$ is the Fermi velocity, $a_{\tb{k}} (b_{\tb{k}})$ denote the Fourier-transforms of $a_j (b_j)$, and the $x$ and $y$ axes are chosen parallel to the nearest neighbor bonds of the $\mathbb{A}$ sublattice. An expansion around $\tb{K}'$ yields a similar Hamiltonian with $k_x$ and $k_y$ interchanged.

To implement interactions, we consider now a balanced mixture of fermionic atoms in two different sub-states (e.g., hyperfine states or Zeeman levels), which play the role of spin-$1/2$ degree of freedom for ordinary electrons. The resulting fermionic on-site Hubbard interaction with strength $U_{FF}$ is tunable via the technique of Feshbach resonances \cite{Tiesinga:93}, which has been recently exploited to realize a Mott state \cite{Ess:08}. However, an extended Hubbard term cannot be easily implemented within the tight-binding approximation. We therefore consider an additional bosonic component coupling to each fermionic component with a strength characterized by $U_{BF}$. The microscopic Hamiltonian of this Bose-Fermi mixture in the tight-binding regime is then given by
\beq
H=H_{0,F}+H_{0,B}+H_{FF}+H_{BB}+H_{BF}.
\eeq
The first two terms are, respectively, given by the kinetic Hamiltonian (\ref{Ham1}) with the annihilation operators $a_i$ replaced by $a_{i,\s}$ for describing fermions with spin $\s$, or by $a_{i,B}$ for the bosons (similarly for the sublattice $\mathbb{B}$). The number operator at site $i$ is defined by $n_{i}=a^\dag_i a_i \,(b^\dag_i b_i)$ for $i\in\mathbb{A} \,(\mathbb{B})$. The Hubbard terms are given by $H_{FF}=(U_{FF}/2) \sum_{i,\s}n_{i,\s}n_{i,-\s}$, which denotes the interaction between fermionic atoms with opposite spins, $H_{BB}=(U_{BB}/2)\sum_{i}n_{i,B}(n_{i,B}-1)$ for the interaction between bosons with $U_{BB}>0$, and $H_{BF}=U_{BF} \sum_{i, \s} n_{i,\s} n_{i,B}$ for the interspecies interaction (the summation over sites in sublattices $\mathbb{A}$ and $\mathbb{B}$ is implicit). The phase diagram for mixtures of bosons with ordinary fermions in an optical lattice is known to comprise a rich phase diagram, with regimes of superfluidity, supersolidity, and phase separation \cite{Bla:03, Demler:05}. 

To derive an effective model, we consider the experimentally well-established regime for the bosons, where they form a superfluid Bose-Einstein condensate with condensate fraction $\tilde{n}$ and healing length $\xi\equiv (t/\tilde{n} U_{BB})^{1/2}$ \cite{Lim:08}. The bosonic sector is accurately described by the Bogoliubov theory, such that the bosonic density-density response function $\chi$ can be computed \cite{Stoof:00, Viverit:00}. Within the parameter regime relevant for experimentally available {\it lattice} Bose-Fermi mixtures (e.g., the widely used rubidium-potassium system) the consideration of the static limit of $\chi$ is sufficient \cite{Illuminati:04}. To obtain the effective interactions for fermions, we note that the phonon-mediated mechanism generates, in general, non-local attractive interaction terms between fermions of all spin states, which fall off on the scale of the healing length $\xi$. Experiments in a bosonic 2D lattice of rubidium atoms by Spielman \textit{et al.} \cite{Spielman:08} show that typical values of $\xi$ are on the order of $d/\sqrt{2}$. With these considerations we arrive at the Hamiltonian $H=H_{0,F}+H_{int,F}$ with
\beq\label{Ham5}
H_{int,F}=-\f{g_0}{2}\sum_{i,\s}n_{i,\s}n_{i,-\s}-g_1\sum_{<ij>}\sum_{\s,\s'}n_{i,\s}n_{j,\s'}.
\eeq
Here, $g_0=(U_{BF}^2/U_{BB})\,\chi_0(\xi)-U_{FF}$ and $g_1=(U_{BF}^2/U_{BB})\,\chi_1(\xi)$ are the on-site and nearest-neighbor interaction strengths, respectively. The static response functions can be readily evaluated numerically to yield $\chi_0(\xi), \chi_1(\xi)\approx 0.2$, whereas the next-nearest-neighbor interactions are one order of magnitude smaller and can thus be neglected. Notice that the value of $U_{BF}^2/U_{BB}$ and $U_{FF}$ may be tuned separately, which allows for an independent control of both $g_0$ and $g_1$.

\begin{figure}
\includegraphics[scale=0.16, angle=0, origin=c]{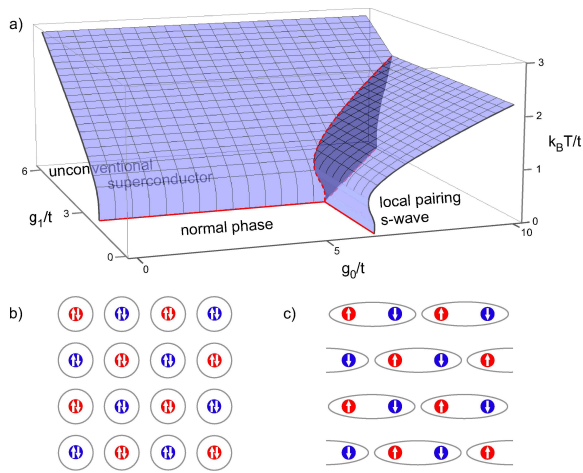}
\caption{\label{Fig.2} (color online) a) Mean-field phase diagram for the Hamiltonian (\ref{Ham5}) with zero chemical potential. b) Real-space configuration of the local pairing $s$-wave superconducting phase.  c) Real-space configuration of the nearest-neighbor spin-singlet bonding with one non-vanishing component. }
\end{figure}
\textit{Competing Superconducting Phases}: We now consider the superconducting (SC) instabilities that the 2D fermions in presence of the staggered gauge field may undergo as a result of the competition between the attractive on-site and the nearest-neighbor interactions ($g_0, g_1\!\!>\!\!0$). While the on-site attraction favors a local pairing $s$-wave spin-singlet SC order, $\langle a_{i \downarrow} a_{i \uparrow} \rangle$ and $\langle b_{i \downarrow} b_{i \uparrow} \rangle$ (see Fig.~\ref{Fig.2}b), nearest-neighbor attractions can lead to an inter-site bond ordering $\langle a_{i\da}b_{i+e_l,\ua}-a_{i,\ua}b_{i+e_l,\da}\rangle$ involving one of the four nearest-neighbor sites (see Fig.~\ref{Fig.2}c), which is called resonating valence bond \cite{Anderson:87}. The vectors $e_l$ with $l \in \{1,2,3,4 \}$ connect adjacent sites of the $\mathbb{A}$ and the $\mathbb{B}$ sublattices, i.e., $e_1=-e_3= (\hat x-\hat y) d/2$ and $-e_2=e_4= (\hat x+\hat y) d/2$. Within BCS mean-field theory, the mean-field Hamiltonian can be written as
\beq
H_{\tr{BCS}}=E_0+\sum_{\tb{k}\in 1BZ} \Psi^\dag_{\tb{k}} \mathcal{D}_{\tb{k}}  \Psi_{\tb{k}},
\eeq
with $\Psi^\dag_{\tb{k}}=(a_{\tb{k},\ua}^\dag, b_{\tb{k},\ua}^\dag,b_{-\tb{k},\da},a_{-\tb{k},\da})$ and
\beq
\label{Matrix}
\mathcal{D}_{\tb{k}}=
\bem
-\mu&-\e^*_{\tb{k}}&\D_{1,\tb{k}}&\D_0\\
-\e_{\tb{k}}&-\mu&\D_0&\D_{1,-\tb{k}}\\
\D_{1,\tb{k}}^*&\D_0^*&\mu&\e_{-\tb{k}}^*\\
\D_0^*&\D_{1,-\tb{k}}^*&\e_{-\tb{k}}&\mu
\eem.
\eeq
Here, $\e_{\tb{k}} =4t[\cos(\phi/4)\cos(k_x d/2)\cos(k_y d/2)-i\sin(\phi/4) \sin(k_x d/2) \sin(k_y d/2)] $ is the full lattice dispersion, $E_0$ is the condensation energy, $\mu$ is the fermionic chemical potential, and the two competing SC orders are expressed as $\D_0, \D_{1,\tb{k}}$ in $k$-space. The free energy is then given by $F(\D_0,\D_{1,\tb{k}})=E_0-k_B T\sum_{n,\tb{k}}\ln \lt[1+\exp(- E_{n,\tb{k}}/k_BT)\rt]$ where $E_{n,\tb{k}}$ with $n \in \{1,2,3,4\}$ denote the four branches of the quasi-particle spectrum of the mean-field Hamiltonian, which are readily obtained by diagonalizing $\mathcal{D}_{\tb{k}}$ in Eq.\ (\ref{Matrix}). The variational parameters  $\D_0$ and $ \D_{1,\tb{k}}$ are self-consistently determined by minimizing the free energy. We remark that the mean-field analysis is justified in the weak coupling regime where the interaction strength is smaller than the bandwidth $g_0/t$, $g_1/t<4 \sqrt{2}$.

The resulting phase diagram at zero chemical potential and $\pi$-flux, shown in Fig.~\ref{Fig.2}a, comprises three phases: an unpaired phase (normal phase), an $s$-wave SC phase and a non-local pairing SC phase.  The SC phases are separated by a first-order phase transition (dark surface between dashed lines), whereas the transition to the normal phase is second order. Due to the vanishing of the density of states at zero chemical potential, the Hamiltonian (\ref{Ham5}) is quantum critical \cite{Neto:07, Annica:07}, i.e. finite critical coupling strengths are required to induce the respective BCS instabilities even at zero temperature. Note that for typical values of the hopping strength $t\sim E_{\tr{R}}$, where $E_{\tr{R}}$ is the recoil energy, the temperature necessary to access the SC phases $k_B T/t\sim \mathcal{O}(1)$ is on the order of $0.1 \,T_F$, which is well in reach of present experiments with ultracold atoms.

The local pairing $s$-wave SC phase is characterized by a gap function $\D_0$ which is constant, whereas the order parameter of the non-local pairing SC phase is a coherent mixture of orbital wave functions with a $p$-wave (odd) and a $d$-wave (even) symmetry,
\beq
\D_{1,\tb{k}} =Ê\D_1 e^{i(k_x+k_y)d/2} \approx \D_1\lt[1+i (k_x +k_y) \frac{d}{2} \rt].
\eeq
The latter approximation holds around the Dirac point $\tb{K}$ at long wavelengths. Due to the lack of inversion symmetry (under interchange of the two sublattices), the order parameter describing the non-local pairing does not have a definite parity. Furthermore, the non-local pairing SC phase in real-space is described by a single bond formed with one of the four nearest-neighbor sites. This results in an unconventional superconductivity: not only is the $U(1)$ gauge group broken in the ordered phase, but also the symmetry group $C_{4v}$ of the underlying crystal lattice.

\begin{figure}
\includegraphics[scale=0.5, angle=0, origin=c]{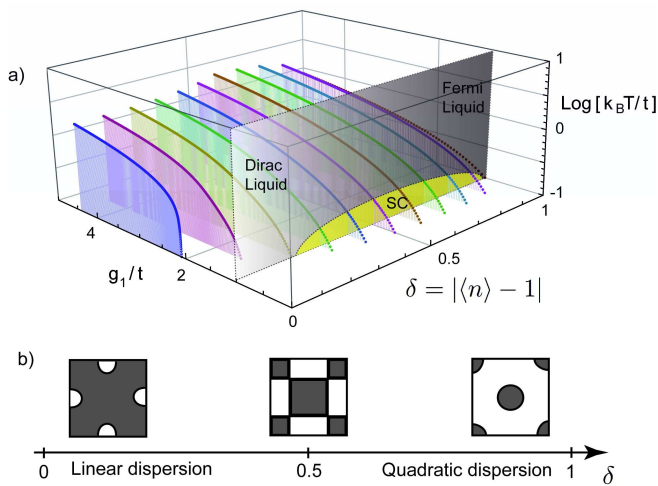}
\caption{\label{Fig.3}(color online) a) $T_c$ in the unconventional SC channel is plotted versus $g_1$ for different fermion filling fractions. At unit filling ($\d=0$) the system is quantum critical, i.e., below a critical value of $g_1$ the critical temperature is zero. Away from unit filling ($\d>0$), $T_c(g_1)$ possesses an exponentially small non-zero tail as $g_1$ approaches zero, which can not be seen in the present scale. The shaded plane at a fixed coupling $g_1/t=1$ shows a dome-like shape of the SC phase embedded in the normal phase. b) Evolution of the non-interacting Fermi surface as a function of doping. The shaded region is the filled Fermi sea.}
\end{figure}

In the following, we consider the case of non-zero chemical potential, focusing on the unconventional SC channel, which yields the highest transition temperature $T_c$. Close to the transition, the fermion filling fraction $\langle n \rangle$ (or the particle/hole doping $\d=|\langle n \rangle-1|$) is trivially determined by the non-interacting limit (ignoring the Hartree energy) as
\beq
\d=\f{1}{2N}\sum_{\tb{k}} \tanh\lt(\f{|\e_{\tb{k}}|+|\mu|}{2 k_B T_c}\rt)-\tanh\lt(\f{|\e_{\tb{k}}|-|\mu|}{2 k_B T_c}\rt),
\eeq
where $N$ is the total number of sites. In Fig.~\ref{Fig.3}a we plotted $T_c$ versus $g_1$ and $\delta$ for the unconventional SC channel. While the dispersion remains linear (Dirac-like) as the filling fraction $\langle n \rangle$ is slightly tuned away from unity ($\mu=0$, $\delta=0$), the system ceases to be quantum critical. The $T_c$ versus $g_1$ graph develops an exponential tail extending all the way to $g_1=0$ (not visible for the scale used in Fig.~\ref{Fig.3}a) and shifts towards lower values of the coupling strength $g_1$. This shift, however, does not grow monotonously as the filling fraction is further decreased but reduces again, below a filling fraction of approximately 0.3 ($\delta \approx 0.7$ in Fig.~\ref{Fig.3}a). Plotting temperature versus doping for a fixed value of the coupling strength $g_1$, as shown in the shaded plane in Fig.~\ref{Fig.3}a, we recognize a dome-shaped unconventional SC phase at intermediate filling fractions, surrounded by the normal phase for fillings close to unity or zero, which we termed Dirac-liquid (Fermi-liquid) on the left (right) side, where linear (quadratic) single-particle dispersion prevails. The dome-structure of the unconventional SC-phase is reminiscent of the phase diagram for high-$T_c$ superconductors. The evolution of the Fermi surface, as we tune the filling fraction from unity to zero, is illustrated in Fig.~\ref{Fig.3}b in the non-interacting limit. The transition between the Dirac-liquid and the Fermi-liquid in presence of interactions is an interesting topic for future studies, which might reveal further correspondences with strongly correlated electron systems. The investigation of a competing charge-density-wave instability is another important task deferred to further studies.

\textit{Experimental Considerations}: A promising experimental realization of the physics described here is provided by a rubidium-potassium system composed of a balanced mixture of fermionic $^{40}$K-atoms prepared in the $|F=9/2,m_F=-7/2\rangle$ and $|F=9/2,m_F=-9/2\rangle$ Zeeman components of the $F=9/2$ ground state hyperfine level and bosonic $^{87}$Rb atoms in the $|F=1,m_F=1\rangle$ ground state \cite{RbK_system:1}. The parameter $U_{BF}^2/U_{BB}$ may be adjusted via its dependence on the well depth, while an $s$-wave Feshbach resonance around 202 Gauss can be used to tune $U_{FF}$ independently. In order to experimentally discriminate the two SC phases discussed here, one could search for a signature of their distinct gap functions in their momentum spectra \cite{Ket:06}. Correlation measurements similar to the one described in Ref.~\cite{Jin:05} should be another powerful method to obtain information on the nature of the pairing.

\textit{Conclusions}:
An ultracold Bose-Fermi mixture in a 2D square optical lattice, subjected to an effective staggered magnetic field, gives rise to strongly interacting Dirac fermions. A rich phase diagram arises, which exhibits a competition between a local pairing $s$-wave and an unconventional superconducting phase at temperatures well in reach of present experimental techniques. The appearance of a dome-like structure of the unconventional superconducting phase, when plotting the critical temperature versus doping for a fixed value of the interaction, reveals an intriguing link to strongly correlated electronic materials. In fact, the staggered $\pi$-flux Hamiltonian studied here is nothing but the mean-field Hamiltonian of Affleck and Marston proposed to describe the pseudogap phase of high-$T_c$ cuprates \cite{Marston:89}. This indicates that our optical lattice model may 
play a prominent role in understanding high-$T_c$ superconductors. The possibility of well-controlled experimental studies of our optical lattice model, including the transition between the Dirac and the Fermi liquid, may reveal further correspondences with high-$T_c$ cuprates or permit to identify missing relevant building blocks yet to be included in the model. 

\begin{acknowledgments} This work was partially supported by the Netherlands Organization for Scientific Research (NWO). AH acknowledges support by DFG (He2334/10-1). We are grateful to D. Baeriswyl, A. H. Castro Neto, K. le Hur, C. Mudry, H. T. C. Stoof, B. Uchoa, M. A. H. Vozmediano, and J. Zaanen for fruitful discussions.
\end{acknowledgments}


\begin{thebibliography}{18}
\bibitem{Neto:09}
A. H. Castro Neto \textit{et al.}, Rev. Mod. Phys. {\bf 81}, 109 (2009).
\bibitem{Jaksch:98}
D. Jaksch \textit{et al.}, Phys. Rev. Lett. {\bf 81}, 3108 (1998)
\bibitem{Greiner:02}
M. Greiner \textit{et al.}, Nature (London) {\bf 415}, 39 (2002).
\bibitem{Spielman:08}
I. B. Spielman \textit{et al.}, Phys. Rev. Lett. {\bf 100}, 120402 (2008).
\bibitem{Hem:07}
A. Hemmerich and C. Morais Smith, Phys. Rev. Lett. {\bf 99}, 113002 (2007).
\bibitem{Lim:08}
L.-K. Lim \textit{et al.}, Phys. Rev. Lett. {\bf 100}, 130402 (2008).
\bibitem{Liu:09}
J.-M. Hou \textit{et al.}, Phys. Rev. A {\bf 79}, 043621 (2009).
\bibitem{Goldman:09}
N. Goldman \textit{et al.}, Phys. Rev. Lett. {\bf 103}, 035301 (2009).
\bibitem{Tiesinga:93}
E. Tiesinga \textit{et al.}, Phys. Rev. A {\bf 47}, 4114 (1993); S. Inouye \textit{et al.}, Nature {\bf 392}, 151 (1998).
\bibitem{Ess:08}
R. J\"{o}rdens \textit{et al.}, Nature {\bf 455}, 204 (2008).
\bibitem{Bla:03}
H. P. B\"{u}chler and G. Blatter, Phys. Rev. Lett. {\bf 91}, 130404 (2003).
\bibitem{Demler:05}
D.-W. Wang \textit{et al.}, Phys. Rev. A {\bf 72}, 051604 (2005).
\bibitem{Stoof:00}
M. J. Bijlsma \textit{et al.}, Phys. Rev. A {\bf 61}, 053601 (2000).
\bibitem{Viverit:00}
H. Heiselberg \textit{et al.}, Phys. Rev. Lett. {\bf 85}, 2418 (2000).
\bibitem{Illuminati:04} 
F. Illuminati and A. Albus, Phys. Rev. Lett. {\bf 93}, 090406 (2004); A. Albus \textit{et al.}, Phys. Rev. A {\bf 68}, 023606 (2003).
\bibitem{Anderson:87}
P. W. Anderson, Science {\bf 235}, 1196 (1987).
\bibitem{Neto:07}
B. Uchoa and A. H. Castro Neto, Phys. Rev. Lett. {\bf 98}, 146801 (2007).
\bibitem{Annica:07}
A. M. Black-Schaffer and S. Doniach, Phys. Rev. B {\bf 75}, 134512 (2007).
\bibitem{RbK_system:1}
S. Inouye \textit{et al.}, Phys. Rev. Lett. {\bf 93}, 183201 (2004); S. Ospelkaus \textit{et al.}, Phys. Rev. Lett. {\bf 97}, 120403 (2006).
\bibitem{Ket:06}
J. K. Chin \textit{et al.}, Nature {\bf 443}, 961 (2006).
\bibitem{Jin:05}
M. Greiner \textit{et al.}, Phys. Rev. Lett. {\bf 94}, 110401 (2005).
\bibitem{Marston:89}
J. B. Marston and I. Affleck, Phys. Rev. B {\bf 39}, 11538 (1989).
\end{thebibliography}
\end{document}